# DA GAUSS A EINSTEIN

RIASSUNTO

Il presente breve saggio, di natura essenzialmente storica, offre un profilo della transizione dalla geometria spazio-temporale euclidea-newtoniana della fisica classica a quella pseudoriemanniana della relatività generale, passando per la geometria pseudoeuclidea della relatività speciale come tappa intermedia.

ABSTRACT

The present short essay "From Gauss to Einstein", of essentially historical nature, aims at describing the transition from the Euclidean-Newtonian space-time geometry of Classical Physics to the Pseudoriemannian geometry of General Relativity, including the Minkowskian geometry of Special Relativity as an intermediate step. (in Italian).

## 1)    VERSO LE GEOMETRIE NON EUCLIDEE

Uscendo da una lunga gestazione, all'inizio del secolo XIX le idee che da oltre due millenni erano a fondamento della geometria cominciarono ad aprirsi verso nuove prospettive. Gli sviluppi ai quali ci riferiamo non avevano tuttavia alcuna relazione con progressi conoscitivi intorno alla natura dello *spazio fisico*, perché nulla era cambiato, e nulla sarebbe cambiato per molto tempo ancora, nella base empirica della disciplina: essi nacquero, ed in quello rimasero fino all'alba del nuovo secolo, in un ambito specificamente teoretico-speculativo. La scoperta dell'esistenza "logica" di geometrie diverse dalla geometria euclidea, iniziata con la rielaborazione critica dell'opera di Euclide da parte di Legendre, e dovuta essenzialmente a Gauss, a J. Bolyai e a Lobatchewsky, [1] appartiene insomma quasi per intero alla matematica.

---

[1] All'amico e compagno di studi W. Bolyai che gli aveva comunicato le scoperte del figlio Janos, Gauss rispose di aver cominciato ad occuparsi dello stesso problema, giungendo a conclusioni simili, «sin da trenta o trentacinque anni prima». Gauss non aveva tuttavia pubblicato niente sull'argomento, forse per evitare le reazioni degli ambienti matematici più retrivi (le famose «strida dei beoti» di cui in una lettera a Bessel. Sembra insomma che la scoperta delle geometrie non euclidee possa essere condivisa tra Gauss, J. Bolyai e Lobatchewsky, il quale ultimo aveva scritto una

Non inaspettatamente, il cardine tra l'idea tradizionale di un'*unica* geometria come possibile "scienza dello spazio", e la nuova visione di una *pluralità* di geometrie dotate dello stesso status logico-matematico, si rivelò risiedere nell'ultimo degli assiomi euclidei: il celebre "quinto postulato" (o "delle parallele"), equivalente all'asserto che una generica retta ha esattamente una parallela passante per un punto dato fuori di essa. A conferma di una lunghissima storia di inutili tentativi di dimostrarne la derivabilità dagli altri quattro, questo assioma fu infine riconosciuto da essi indipendente; sostituendolo quindi con una sua conveniente diversa versione (ad esempio con l'affermare l'esistenza di *più* di una parallela per quel punto), diventava possibile affacciarsi all'orizzonte di una nuova geometria, avente gli stessi "diritti di cittadinanza", sul piano logico, della geometria euclidea classica.

Per le stesse ragioni ascrivibile soltanto alla matematica fu un risultato di natura tecnica pubblicato da Gauss nel 1828 assieme ad altri importanti sviluppi dell'allora nascente geometria differenziale delle superfici immerse nello spazio euclideo tridimensionale. [2] Gauss considera una porzione $M^2 \equiv M$ di superficie immersa, data localmente in forma parametrica assegnando le tre coordinate cartesiane del suo generico punto P come funzioni abbastanza regolari di due parametri (o coordinate) (u,v), dunque secondo P = P(u,v), in un intorno aperto e semplicemente connesso $U(u_0,v_0) \subset R^2$ di una coppia di riferimento $(u_0,v_0)$ alla quale corrisponde il punto $P_0 =: P_0(u_0,v_0)$. Mediante queste tre funzioni Gauss calcola, a loro volta come funzioni di (u,v), i tre coefficienti E, F, G di una forma quadratica in due indeterminate $(\xi,\eta)$, diciamo $Q(u,v|\xi,\eta) =: E(u,v)\xi^2 + 2F(u,v)\xi\eta + G(u,v)\eta^2$, il cui valore in (u,v) uguagli il quadrato della lunghezza di un breve arco di M passante per P(u,v) quando le indeterminate $(\xi,\eta)$ siano identificate con le piccole differenze $(\Delta u, \Delta v)$ tra le coordinate omologhe degli estremi dell'arco. [3] È evidente che il campo (E,F,G)(u,v) in $U(u_0,v_0)$, e in particolare i valori di E, F, G e delle loro derivate prime e seconde (rispetto a u e/o v) in $(u_0,v_0)$, è riconducibile a misure (di lunghezze e di angoli) interne a M, o come anche si dice, a misure relative alla "geometria intrinseca" di M. Orbene, Gauss trova che il prodotto delle due

---

curvature principali [4] di M in $P_0$ (a prima vista *non riconducibili* alla geometria intrinseca di M) è tuttavia esprimibile come ben definita funzione di E, F, G e loro derivate prime e seconde in $P_0$. In particolare, questa funzione risulta nulla se tali derivate prime e seconde sono nulle, e quindi identicamente nulla in $U(u_0,v_0)$ se E, F, G sono ivi costanti.

La portata di una tale scoperta non poteva certo sfuggire all'attenzione di Gauss, il quale non esitò (forse con una punta di autocompiacimento) a battezzarla con il nome di «Theorema egregium». Per esprimerci con una nota e suggestiva immagine, essa implicava che una ipotetica minuscola creatura "bidimensionale" residente in M, e quindi incapace di percepirne la configurazione nello spazio 3-dimensionale sommergente (ovvero la "geometria estrinseca"), sarebbe stata in grado di risalire al prodotto tra le sue due curvature principali ($\equiv$ curvatura totale), che di quella configurazione sembrava essere espressione, mediante certe misure *interne* a M, o misure intrinseche, a lei accessibili.

Il Theorema egregium ebbe un'influenza decisiva sull'accettazione delle geometrie non euclidee. Ovviamente un piano euclideo (o una sua porzione aperta, analoga alla porzione di superficie considerata da Gauss) ha ovunque, e per qualunque sua sezione normale, curvatura nulla (ciò si esprime comunemente dicendo che un piano è "piatto"). Quindi la nozione di curvatura principale vi degenera, nel senso che le curvature di tutte le sezioni normali possono considerarsi principali alla luce della loro definizione, e sono comunque nulle. Ciò è confermato dal Theorema egregium, in quanto – come avviene appunto nel caso del piano euclideo – esistono coordinate tali che E, F, G risultano costanti. (L'esistenza di tali coordinate è fin qui soltanto sufficiente a che la curvatura totale sia identicamente nulla; ma si dimostra che essa è anche necessaria.)

Si affacciava così l'idea che la correttezza del modello euclideo di un piano fisico (e per immediata estensione dello spazio fisico) potesse diventare oggetto di indagine sperimentale. A quel punto, cioè, il problema era quello *di effettuare certe specifiche misure intrinseche* con la precisione necessaria, per verificare la presunta "piattezza" dello spazio fisico; dopodiché, nella eventuale assenza di una sua conferma, si sarebbe convenientemente modificato qualcuno degli assiomi della geometria euclidea. In particolare sembra certo che Gauss si sia spinto a misurare la somma degli angoli interni di un triangolo avente per vertici tre vette mutuamente visibili delle montagne dello Harz, trovando – come non fa oggi meraviglia tenuto conto della estrema esiguità dell'effetto e della sensibilità degli strumenti dell'epoca – 180° entro gli errori sperimentali. Così il problema restò senza una risposta definitiva; ma almeno dal punto di vista filosofico, il nuovo approccio alla geometria inferse un duro colpo all'apriorismo conoscitivo di matrice kantiana («giudizi sintetici a priori»), e in una lunga prospettiva, contribuì significativamente al tramonto di tutte le metafisiche (o almeno di tutte le commistioni tra la metafisica e la scienza). [5]

---

[4] Ricordiamo che le curvature principali di una superficie regolare S, in un suo punto prefissato, sono i due valori stazionari del reciproco del raggio (con segno) del cerchio osculatore della curva $\Gamma$ sezione di S con un piano $\pi$ normale ad S, al ruotare di $\pi$ attorno al versore normale N ad S stessa. (Il segno del raggio si riferisce all'orientamento di N, prefissato arbitrariamente una volta per tutte.) Sebbene dimensionalmente sia il *quadrato* di una curvatura, il prodotto delle curvature principali si dice tradizionalmente curvatura "totale" (o "gaussiana") di S, ed evidentemente il suo segno *non* dipende dall'orientamento di N.

Il giovane Riemann dovette essere molto colpito dalle scoperte geometrico-differenziali del suo celebre tutor quando, intorno alla metà del secolo, cominciò ad interessarsi ai fondamenti della geometria come teoria fisico-matematica; al punto che, nella sua famosa dissertazione del 1854 tenuta a Göttinga [6] su questo tema, propose di sviluppare la geometria – non soltanto delle superficie, ma di quelle che più tardi si sarebbero chiamate "varietà n-dimensionali" – in modo completamente *intrinseco*. Chiariamo meglio i termini della questione. Come abbiamo visto, Gauss muoveva innanzitutto dalla considerazione di una data porzione aperta di superficie immersa nel normale spazio euclideo 3-dimensionale, e calcolava i tre coefficienti E, F, G della forma Q (definita positiva) come funzioni delle due coordinate (u,v). Pur rendendosi conto che così facendo apriva le porte ad una geometria intrinseca di quella (porzione di) superficie, autonoma rispetto alla sua immersione nello spazio 3-dimensionale, almeno in apparenza egli non trasse tutte le conseguenze di questa circostanza. Invece Riemann radicalizzò l'idea di geometria intrinseca *partendo* dalla nozione di forma Q (o metrica) di una varietà n-dimensionale come un dato, ossia pensandone gli $n(n+1)/2$ coefficienti (3 nel caso $n = 2$ della superficie, le soprannominate funzioni

---

[5] Proprio Gauss fu fiero oppositore di quei filosofi a lui contemporanei che avevano la pretesa di pontificare sulle scienze esatte nonostante la loro insufficiente o nessuna competenza in materia. In una lettera del 1844, egli cita in proposito Shelling, Hegel e von Essenbeck, aggiungendo: «E anche con lo stesso Kant, spesso non va molto meglio: secondo me, la sua distinzione tra proposizioni sintetiche e analitiche è una di quelle cose che o cadono nella banalità o sono false». A quel tempo Gauss aveva già una conoscenza matura della geometria non euclidea, la quale è di per sé una netta confutazione delle idee di Kant in materia di spazio e relativi giudizi sintetici a priori. Non possiamo passare sotto silenzio, allo stesso proposito, la parola di A. Einstein: «Sono convinto che i filosofi hanno sempre avuto un effetto nefasto sul progresso del pensiero scientifico, poiché hanno sottratto molti concetti fondamentali al dominio dell'empirismo, nel quale si trovavano sotto il nostro (dei fisici, ndr) controllo, trasferendoli alle intangibili altezze dell'*a priori* (evidente allusione a Kant, ndr). Infatti, anche se dovesse risultare che il mondo delle idee non può essere dedotto dall'esperienza attraverso mezzi logici ma è, in un certo senso, una creazione della mente umana, senza la quale non è possibile nessuna scienza, il mondo delle idee risulterebbe altrettanto indipendente dalla natura delle nostre esperienze quanto lo sono i vestiti dalla forma del corpo umano.» (da "The Meaning of Relativity", 5ª ed. Princeton Un. Press 1953). Il coinvolgimento di molti filosofi in campi conoscitivi che, pur competendo a pieno titolo alla filosofia, risultano al di là delle loro effettive conoscenze specifiche, è fenomeno duro a morire. Per citare soltanto due esempi estremi, possiamo ancora ricordare la manifesta incapacità di comprendere la teoria della *relatività speciale* da parte di H. Bergson ("Durée e simultanéité: à propos de la théorie d'Einstein" (1922). «Dio lo perdoni» fu il laconico commento del "chiamato in causa"); e molto più recente, la dissacrante monografia-beffa di Sokal e Bricmont ("Impostures intellectuelles", Éditions Odile Jacob (1997)), vera e propria "galleria degli orrori" di certa filosofia sedicente scientifica del Novecento, specialmente francese. Generalmente parlando, da circa due secoli gli scienziati rimproverano ai filosofi che scelgono la scienza come oggetto delle loro meditazioni la troppo frequente mancanza di rigore, di concretezza e addirittura di informazione; ciò che a lungo andare ha prodotto una significativa dose di indifferenza, quando non di franca *insofferenza*, da parte del mondo scientifico nei confronti di quello filosofico.

[6] B. Riemann, "Über die Hypothesen, welche der Geometrie zu Grunde liegen", in Abh. der Ges. zu Gött., XIII (1868); anche in Werke, 2ª ed., p. 272; vedi anche in D.E. Smith, "A Source Book in Mathematics" (1929), ristampa in Dover (1959), p. 411. Una forse migliore traduzione inglese del testo riemanniano si trova nel 2° volume del trattato di M. Spivak "Differential Geometry", 5 vols, curata dallo stesso autore. Infine una traduzione italiana (dovuta a G. Gabella) è per la prima volta accessibile al grande pubblico in "La grande biblioteca della Scienza", Fabbri Editori 2008.

E, F, G) come funzioni convenientemente regolari delle n coordinate, date sotto la sola condizione (di non-degenerazione) che il determinante della relativa (n×n)-matrice simmetrica (cioè EG − F² per n = 2) non fosse nullo nel dominio di interesse. A priori, questa metrica poteva essere definita positiva (come nel caso delle varietà n-dimensionali immerse in uno spazio euclideo (m≥n)-dimensionale, quando tutti i minori principali della matrice dei coefficienti, incluso il suo stesso determinante, sono positivi), o come si cominciò presto a dire, "riemanniana"; ma anche, più generalmente, indefinita o "pseudoriemanniana".[7]

Anche se sotto certi aspetti lontana dal moderno standard di rigore, la riformulazione riemanniana fu quanto di più radicalmente innovativo fosse stato proposto fino ad allora sui fondamenti della geometria. Essa detronizzò la geometria euclidea, che fu ridotta, per così dire, al rango di mera "geometria possibile" (allora, dal solo punto di vista matematico), e conferì pieno status logico alla geometria iperbolica di J. Bolyai e Lobatchewsky, che Riemann probabilmente non conosceva, e che quegli autori avevano formulato sostituendo le usuali funzioni circolari con quelle iperboliche nella geometria della sfera.[8] Ma soprattutto, il lavoro di Riemann sottolineò che ogni ricerca sulla natura dello spazio *fisico* doveva partire dall'esperienza, distruggendo così definitivamente, almeno nel giudizio delle menti meno ottuse, ogni illusione circa una "conoscenza a priori" del mondo.

Riemann concluse la sua conferenza con queste parole, la cui capacità profetica non può non colpire profondamente: « … o la realtà soggiacente allo spazio costituisce una varietà discreta, o bisognerà ricercare il fondamento delle sue relazioni metriche al di fuori di esso, *nelle forze di connessione che vi operano*. … Ciò conduce *nel dominio della fisica*, nel quale il tema delle presenti ricerche non ci consente di entrare.» (corsivi dr) Una tale allusione a quello che in ultima analisi, e passando dallo "spazio" allo "spazio-tempo", sarà il tema della relatività generale, lascia stupefatti, perché ai tempi in cui fu concepita, ed anche assai più tardi, l'idea che ipotetiche «forze di connessione» (tra corpi materiali?) potessero influenzare la natura *metrica* dello spazio (o addirittura dello spazio-tempo) appariva come una vera e propria fantasticheria.

---

[7] Ricordiamo che nel caso indefinito ci si riferisce alla "pseudolunghezza" del piccolo arco della varietà come alla radice quadrata del *valore assoluto* della forma Q(ξ,η). In particolare possono esistere archi (con estremi distinti) di pseudolunghezza nulla, esattamente come nello spazio pseudoeuclideo possono esistere vettori non nulli con pseudomodulo nullo.

[8] Questa scoperta fu pubblicata soltanto nel 1835 (Lobatchewsky). Fu merito di E. Beltrami quello di aver costruito (1866) una superficie *a curvatura costante negativa* immersa nel normale spazio euclideo (precisamente, la superficie generata dalla rotazione attorno al loro comune asintoto di due archi di trattrice simmetrici rispetto al piano equatoriale, detta anche "pseudosfera"), che risultava essere un modello locale del piano iperbolico di J. Bolyai e Lobatchewsky. Molto più tardi (1901), Hilbert dimostrò che non può esistere una superficie analitica (del 3-spazio euclideo) che sia un modello *globale* del piano iperbolico. Quest'ultimo risultato è significativo nell'ottica della futura cosmologia relativistica, suggerendo (in modo vago, ma ante litteram) la presenza di singolarità nelle soluzioni, prolungate analiticamente in tutto il loro dominio di esistenza, delle equazioni gravitazionali (vedi anche i cenni, verso la fine della prossima sezione, alle soluzioni di Schwarzschild e di Friedmann di quelle equazioni).

2)    LA RELATIVITÀ SPECIALE

La prima e poco avvertita scossa al quadro che fino ad allora aveva rappresentato lo spazio ed il tempo come "contenitori" assoluti e separati degli eventi fisici che vi occorrono, doveva arrivare dalla teoria dell'elettromagnetismo (pubblicata da Maxwell nel 1873). Ben presto, infatti, ci si rese conto che le equazioni ElettroMagnetiche (EM) non erano *esattamente* invarianti a fronte di trasformazioni galileiane, sia delle coordinate che dei campi EM [9] , ma soltanto a meno di termini dell'ordine del prodotto $\varepsilon_o\mu_o \equiv c^{-2}$ delle permeabilità elettrica e magnetica nel vuoto. Tuttavia la comunità scientifica convisse relativamente a lungo con questa difformità, secondo la quale le equazioni della meccanica, ma non quelle EM, risultavano esattamente invarianti rispetto alle trasformazioni galileiane.

Intorno ai fenomeni EM, da una parte dominava l'idea meccanicistica di "campo" come stato fisico di un ipotetico mezzo continuo, ma dall'altra non sembrava possibile identificare questo mezzo con alcunché di materiale. In mancanza di meglio, qualcuno pensò di chiamare l'ipotetico supporto immateriale dei campi EM "etere luminifero". Ma il problema, evidentemente, era quello di conoscere quale era il moto del laboratorio, ove si effettuavano osservazioni e misure, rispetto all'etere.

Il primo, e poi più noto e definitivo, esperimento in tal senso (ripetuto in una lunga serie di versioni sempre più raffinate), fu quello di Michelson [10] (a partire dal 1881, e più tardi in collaborazione con Morley). Esso non rivelò peraltro alcun moto del laboratorio, o della stessa Terra, rispetto all'etere. Esito del pari negativo ebbero altri esperimenti basati su effetti diversi: quello di Rayleigh (1902) [11]  e quello di Brace (1904) [12] , che sfruttavano entrambi la doppia rifrangenza di certe sostanze; e quello di Trouton e Noble (1903) [13] , che misurarono la coppia di torsione agente su un condensatore a placche trasverse alla presunta direzione del moto.

L'accumularsi di queste risposte negative circa l'esistenza di un moto (del laboratorio, della Terra, o dello stesso sistema solare) rispetto all'etere, spostarono presto l'attenzione sulle trasformazioni delle coordinate (x,y,z,t) di sistemi di riferimento in moto relativo rettilineo

---

uniforme, da sostituire, per generalizzarle, alle trasformazioni galileiane. Quasi fatalmente, questo collocò al centro della nuova cinematica una nozione algebrica che risaliva a oltre un secolo prima: precisamente, quella di gruppo delle trasformazioni delle coordinate (di quei sistemi di riferimento). [14] Chi per primo capì che alla radice del problema era appunto l'individuazione del gruppo *corretto* (visto che quello galileiano doveva essere rifiutato), sembra essere stato H. Poincaré.

È in queste circostanze che il "tempo" (o meglio il prodotto del *modulo* della velocità della luce *nel vuoto* c, riconosciuto come una costante assoluta, per il tempo t – la cosiddetta "lunghezza römeriana" ct), da sempre visto come parametro che "etichetta" una successione continua di copie identiche dello spazio fisico (euclideo), fa la sua piuttosto obliqua comparsa sulla scena di una nuova geometria 4-dimensionale. Le trasformazioni di Lorentz, che "mescolando" i due tipi di coordinate (spaziali e temporale) assicurano alle equazioni di Maxwell la desiderata invarianza al passaggio tra riferimenti in moto relativo uniforme, sono pubblicate nel 1904 [15] . (Beninteso, a tal scopo alle trasformazioni di Lorentz vanno associate le trasformazioni *relativistiche* dei campi EM e delle loro sorgenti J, ρ.) Esse precedono quindi di pochissimo il lavoro di Einstein del 1905 [16] con il quale la relatività speciale si afferma quasi subito come conquista *teoretica* definitiva. Come vi era ben da aspettarsi, a questo punto la coperta risulta tuttavia troppo corta, nel senso che viene ora a mancare, rispetto alle nuove trasformazioni, l'invarianza della dinamica newtoniana; ma Einstein non esita a porre rimedio a questa difficoltà *modificando opportunamente quella dinamica*. [17]

La vicenda, assolutamente "fatale" per la geometria fisica e per la fisica in generale, è storicamente troppo importante per non descriverla con qualche dettaglio. Essa culmina in un brevissimo periodo a cavallo tra il 1904 e il 1905, pur avendo il suo diretto punto di avvio nel primo esperimento di Michelson di quasi quindici anni prima, e alcune sue premesse teoretico-speculative nelle idee espresse sin dai primi anni '80 dal fisico-epistemologo-fisiologo E. Mach. Partendo da suoi "criteri di verificabilità" estremamente rigorosi, Mach rigetta innanzitutto le nozioni di spazio e tempo assoluti come "metafisiche". Egli sostiene infatti a ragione che non osserviamo né l'uno né l'altro in sé, ma solo gli *eventi* che vi occorrono; e dunque, che ogni tentativo di separare la geometria fisica dal resto della fisica *va incontro a insormontabili difficoltà logiche*, a vere e proprie aporie. In particolare, ma questo riguarda già la relatività generale, Mach vede l'inerzia di un corpo materiale come una sua relazione con tutta la materia dell'universo (il cosiddetto "principio di Mach"). Queste idee di carattere fondamentale avranno una non trascurabile influenza su Einstein durante il suo laborioso sviluppo della teoria della relatività generale. [18]

---

[14] Si tenga presente che se l'operazione del gruppo è una dipendenza funzionale (come nel caso delle trasformazioni delle coordinate), l'associatività è automaticamente assicurata dalla definizione di funzione.
[15] H.A. Lorentz, "Electromagnetic Phenomena in a System moving with any Velocity less than that of Light", Proc. Acad. Sci. Amsterdam, 6 (1904).
[16] A. Einstein,  "Zur Elektrodynamik bewegter Körper", Annalen d. Physik **17**, 891 (1905).
[17] Torna qui opportuno porre in evidenza la "cifra" dell'approccio alla filosofia naturale di Einstein, e cioè la sua natura spregiudicatamente pragmatica, con l'apprezzamento della capacità predittiva delle teorie come loro qualità decisiva. Ben nota in proposito è del resto l'opinione (dello stesso Einstein) secondo la quale «il vero scienziato può apparire all'epistemologo sistematico come un opportunista privo di scrupoli.». D'altra parte non va dimenticato che le modifiche apportate da Einstein alla dinamica del punto materiale erano già state proposte da Lorentz nella memoria del 1904, § 10.

La storia del progresso teorico verso il traguardo einsteiniano del 1905 può farsi cominciare nel 1892, quando FitzGerald [19] propone di giustificare l'esito negativo dell'esperimento di Michelson con una congettura ad hoc a prima vista un po' bizzarra, e secondo la quale un corpo in moto con velocità v (rispetto all'ipotetico etere) di modulo < c, subisce una contrazione, nella direzione di v, per un fattore $(g(|v|))^{-1} =: (1 - v^2/c^2)^{1/2} \equiv 1/\beta < 1$. L'ipotesi di questo effetto è avanzata poco tempo dopo, ma indipendentemente, da Lorentz [20], e da allora l'effetto stesso si chiamerà "contrazione di Lorentz-FitzGerald". Ancora a Lorentz si deve l'introduzione del "tempo locale" $\tau =: g(|v|)(T - vX/c^2)$ indicato da un orologio in moto uniforme con velocità v (sempre di modulo < c) lungo l'asse X del riferimento "fisso" ($\equiv$ solidale con l'etere), e transitante dall'origine (X = 0) al tempo T = 0. Sostituendo X con vT nella precedente definizione di $\tau$, risulta $\tau = (g(|v|))^{-1}T$; ovvero il tempo di Lorentz, indicato dall'orologio in moto, *ritarda* rispetto a quello dell'orologio fisso.

Nel suo lavoro del 1904, Lorentz svolge una sistematica indagine su come si debbano trasformare gli operatori ($\nabla, \partial/\partial t$), e i campi E (elettrico), H (magnetico), $\rho$ (di densità di carica) e u (di velocità della carica), al passaggio da un riferimento "fisso" $\mathfrak{S}$ ad un riferimento $\mathfrak{s}$ in moto uniforme rispetto al primo con velocità $\Upsilon$ di modulo < c, diciamo lungo un comune asse X $\equiv$ x, affinché le equazioni di Maxwell nello spazio *vuoto* (vuoto a parte la presenza delle cariche), contenenti quegli operatori e campi, si mantengano formalmente invariate trattandovi c come una costante assoluta [21]. Dovrebbe tuttavia esser chiaro che, così formulato, il problema che Lorentz si pone non può avere una risposta univoca; ce l'ha, semmai, quello di determinare la legge di trasformazione dei campi (E,H,$\rho$,u) a partire da quella degli operatori ($\nabla, \partial/\partial t$), o il viceversa. [22] Il

---

[18] Einstein ebbe ammirazione per Mach – in particolare per la sua idea circa l'origine dell'inerzia –, ma prese più tardi posizione contro il di lui approccio filosofico alla conoscenza del mondo fisico. «Il sistema di Mach – ebbe a dire in una conferenza nel 1922 – studia le relazioni esistenti tra i dati sperimentali; secondo Mach, la scienza è la totalità di queste relazioni. Si tratta di un punto di vista scorretto: in effetti quello di Mach è un catalogo, non un sistema.» Evidentemente, Einstein poneva al vertice della conoscenza del mondo fisico ciò che legava le sopraddette relazioni, e che faceva del loro catalogo un sistema. Se fosse stato un logico, avrebbe espresso la sua richiesta dicendo che le relazioni tra i dati sperimentali dovevano essere in corrispondenza biunivoca con le "relazioni" di un sistema formale.
[19] Vedi in O. Lodge, London Transact. (A) **184**, 727 (1893).
[20] H.A. Lorentz, Amst. Verh. Akad. v. Wet. **1**, 74 (1892). Si veda anche, dello stesso autore, "Versuch einer Theorie der elektrischen und optischen Erscheinungen in bewegten Körpern", Brill, Leiden (1895).
[21] Il valore di c è ormai noto con precisione non inferiore a $10^{-10}$; espresso con 6 cifre significative, è uguale a 2,99793 $10^8$ m/sec.
[22] Immaginando di trasportare lo stesso problema sull'equazione della dinamica newtoniana per un punto di massa m > 0 e soggetto ad una forza f, sarebbe come pretendere di derivare dall'equazione stessa, richiedendone l'invarianza

fatto che egli riesca nondimeno ad individuare correttamente *entrambe* le leggi (a parte una secondaria inesattezza relativa alle trasformazione delle sorgenti ρ e ρu), e per giunta in assenza di una ragionevole e dichiarata base di *principi fisici*, va pertanto attribuito ad un insieme di acute e fortunate intuizioni. Naturalmente le trasformazioni di Lorentz comportano sia l'effetto di contrazione (di Lorentz-FitzGerald) che la nozione di tempo locale (di Lorentz).

Pochissimo tempo dopo, Poincaré fornisce le corrette trasformazioni delle sorgenti, ottenendo così un quadro completo e autoconsistente, ma ancora insoddisfacentemente fondato dal punto di vista fisico, dell'elettromagnetismo nel vuoto. Combinando questi risultati con l'espressione della forza di Lorentz su una carica elettrica (invariante) unitaria "test" (cioè "di prova"), Poincaré addirittura intravede la prospettiva di una meccanica "diversa" di cui si fa profeta (seppur non troppo convinto, e per brevissimo tempo): «Forse dovremo formulare una meccanica completamente nuova, sulla quale abbiamo per ora soltanto gettato uno sguardo, in cui l'inerzia cresce con la velocità e la velocità della luce è un limite insuperabile.» [23] Come abbiamo già accennato, Poincaré vede anche immediatamente, e sottolinea, il fatto che le nuove trasformazioni condividono con quelle galileiane il carattere gruppale; e propone di intitolarle, come poi sarà, a colui che le aveva per primo introdotte. [24]

A questo punto le trasformazioni di Lorentz sono quanto occorre e basta per costruire senza grandi problemi una "fisica relativistica macroscopica in assenza di gravità" (geometria fisica inclusa) formalmente invariante rispetto alle dette trasformazioni [25]; come farà appunto Einstein (forse ancora all'oscuro dei risultati di Lorentz-Poincaré) nel citato lavoro del 1905, e come del resto aveva in buona parte già portato a termine lo stesso Lorentz nel 1904. È quindi naturale chiedersi perché il contributo einsteiniano abbia quella carica di novità che la storia della Scienza

---

formale rispetto a cambiamenti di riferimenti, *sia* le trasformazioni galileiane *che* l'invarianza del rapporto f/m. Sfortunatamente, l'affermazione che le equazioni di Maxwell sono invarianti rispetto alle trasformazioni di Lorentz" (senza altro aggiungere) ricorre continuamente nella letteratura (didattica e non), e noi stessi ne abbiamo usato una versione indebolita poco più sopra.

[23] H. Poincaré, Conferenza di St. Louis del settembre 1904, pubblicata in Monist, **15**, 1 (1905). Si veda anche, dello stesso autore, "Sur la dynamique de l'électron", C.R. Paris **140**, 1504 (1905). Tutto considerato, Poincaré fu veramente ad un passo dalla relatività speciale: ma apparentemente non la capì nella sua sostanza fisica, ostinandosi fino alla morte (1912) in un imbarazzato (e imbarazzante) agnosticismo. Interessanti riflessioni dello stesso Poincaré sulla natura dello spazio e del tempo si trovano anche nelle sue "Dernières pensées", Cap. 2, (1913).

[24] Per la verità, trasformazioni delle coordinate anche più generali di quelle di Lorentz erano già state considerate da W. Voigt nel 1887 (Gött. Nachr. 45 (1887)), nell'ambito di uno studio sul vecchio "modello elastico" delle vibrazioni luminose. Nel suo noto trattato storico-critico ("Electromagnetics", Longmans (1938), ripubblicato da Dover (1965, 2 vols.), A. O'Rahily trae spunto dalle trasformazioni di Voigt per una lunga discussione (Cap IX, appunto intitolato "Voigt") sul rapporto tra l'elettromagnetismo e la relatività speciale. Il trattato di O'Rahily è certamente pregevole per ricchezza di notizie ed osservazioni critiche, ma le conclusioni che vi appaiono in materia di relatività speciale non potevano essere prese in seria considerazione, con largo margine, già alla data della sua pubblicazione.

[25] A rigore, poiché in tale fisica relativistica speciale compaiono masse/energie, e queste sono comunque associate ad effetti gravitazionali, esse masse/energie, nonché le sorgenti del campo elettromagnetico, andranno supposte abbastanza piccole. È precisamente questo il significato dell'attributo "test" usato nel paragrafo precedente. Alternativamente, si potrà pensare ad un modello "speciale" in cui la costante di Cavendish venga azzerata, o modello cosiddetto "a gravità spenta".

usualmente gli attribuisce. Due sono le ragioni che ne rendono conto (ma a nostro avviso, *soltanto in parte*): vale a dire, (i) la nuova teoria *è fisicamente fondata*, cioè poggia su principi fisici dichiarati e pienamente ragionevoli alla luce dei fatti sperimentali, principi dai quali le trasformazioni di Lorentz *discendono come logica conseguenza*; (ii) anche se ne sono sostanzialmente illuminati, gli sviluppi che da quei principi portano alle trasformazioni di Lorentz sono *indipendenti dalla teoria elettromagnetica*, non considerando la supposta costanza "universale" (del modulo) della velocità della luce nel vuoto c. [26]

Ricordiamo che per definizione un riferimento (x,y,z,t) è inerziale se in esso vale la legge d'inerzia, ovvero se un punto materiale sottratto all'azione di qualunque forza si muove con velocità (vettoriale) costante. Nel caso della dinamica newtoniana, questa è una conseguenza ovvia della equazione fondamentale ma = f, e della invarianza di m e di f. Mantenendo la sopramenzionata definizione di riferimento inerziale, ricordiamo anche che l'insieme dei fondamenti della relatività speciale (inclusi i protocolli metrologici, e in particolare quelli relativi alla sincronizzazione degli orologi locali) è costituito dai principi seguenti: (1) un "principio (o assioma) di relatività" (≡ le leggi fisiche devono mantenere la stessa forma in tutti i riferimenti inerziali, anche se le misure di certe quantità possono dare valori numerici diversi); (2) un "principio della costanza di c" (≡ misurato secondo gli opportuni protocolli, il modulo della velocità della luce nel vuoto è lo stesso per tutti i riferimenti cartesiani inerziali, beninteso con le stesse unità di lunghezza e di tempo [27] ); (3) un "principio di linearità" (≡ le leggi di trasformazione delle coordinate spazio-temporali dei riferimenti inerziali devono essere lineari affini − o lineari se gli eventi con coordinate spazio-temporali nulle nei due riferimenti coincidono −, con coefficienti dipendenti a priori dalla loro velocità relativa); (4) un "principio di simmetria della misura di lunghezze perpendicolari", o in breve di "simmetria perpendicolare" (≡ la lunghezza di un segmento-campione solidale al riferimento mobile e *perpendicolare* alla direzione del moto relativo tra i riferimenti (di velocità $\Upsilon$ con $|\Upsilon| < c$), non dipende dal *verso* di questo moto, cioè dal *segno* di $\Upsilon$). [28 , 29]

Sotto queste condizioni, e nulla di più, Einstein *deduce* le trasformazioni di Lorentz; e inserendo queste ultime nel sistema di Maxwell, di cui richiede l'invarianza formale, similmente

---

[26] Vale la pena di riportare le dichiarazioni "di non-priorità" di Lorentz, in materia di relatività speciale, che figurano in un suo scritto del 1928 (Astr. Journ. **68**, 350 (1928)): «Io consideravo la mia trasformazione del tempo soltanto come un'ipotesi di lavoro, per cui la teoria (della relatività speciale, ndr) è in realtà opera del solo Einstein. E non può esservi dubbio che egli l'avrebbe ugualmente concepita anche se il lavoro di tutti i suoi predecessori in questo campo non fosse esistito. In questo senso, la sua opera è indipendente dalle teorie che l'hanno preceduta».

[27] Qui si presuppone che i regoli siano "rigidi", e similmente che gli orologi siano "normali"; cioè, sia gli uni che gli altri, sottratti ad ogni influenza che ne possa presuntivamente alterare la lunghezza e rispettivamente il ritmo di marcia.

[28] Gli ultimi due principi appaiono ad Einstein talmente ovvi che egli nemmeno si cura di nominarli esplicitamente come tali.

[29] Mirando ad una fondazione assiomatica completa della relatività speciale, gli assiomi (1÷4) soprariportati non ne esauriscono l'elenco. Uno degli assiomi non nominati è ad esempio quello che recita: "esiste un riferimento inerziale".

*deduce* le leggi di trasformazione dei campi (E,H,ρ) (quella relativa a u ≡ campo di velocità della carica disgregata, o di un qualunque punto in moto prescritto, segue direttamente dalle trasformazioni di Lorentz). Segue anche, in particolare, che la carica contenuta in un piccolo volume mobile è invariante; un asserto non nuovo (Poincaré), ma talvolta in precedenza invocato come ipotesi. Assimilando le forze *di qualunque natura* a quella di Lorentz sulla carica unitaria test, Einstein modifica infine la legge dinamica newtoniana mdv/dt = f (invariante rispetto alle trasformazioni di Galileo ma non rispetto a quelle di Lorentz) con l'introdurvi una "massa di moto" (*sotto* il segno d/dt!) in luogo della massa invariante di Newton – uguale a quest'ultima per v = 0 e crescente con il modulo di v secondo il fattore g(|v|) –, e poi assumendo una consistente legge di trasformazione della forza f (non più invariante come nella dinamica newtoniana). [30] Come ci si aspetta, un riferimento inerziale resta tale sotto una trasformazione di Lorentz-Poincaré.

Questi risultati di eccezionale importanza portano ad una sintesi coerente e fisicamente fondata della meccanica e dell'elettromagnetismo in assenza di gravità. La meccanica relativistica così formulata riproduce la meccanica newtoniana quando vi si faccia c → ∞, e quindi le trasformazioni di Lorentz degenerino in quelle di Galileo. Nel giro di pochi mesi, la "nuova meccanica" intravista da Poincaré diventa una realtà teorica autoconsistente, pronta ad affrontare le necessarie convalide sperimentali.

Le trasformazioni di Lorentz saranno presto dedotte anche da altri sistemi di assiomi [31]; uno di questi assiomi, che aveva già ricevuto notevole interesse matematico nell'ambito dell'algebra degli spazi lineari con metrica generalmente *indefinita*, è quello dell'invarianza (rispetto al solito passaggio tra riferimenti inerziali) della metrica pseudopitagorica $\Delta l^2 - c^2 \Delta t^2$, dove $\Delta l$ è la separazione spaziale, e $\Delta t$ quella temporale, tra due eventi qualsiasi (qui $\Delta l^2$ sta per $(\Delta l)^2$, et sim.). In particolare, questa algebra apre la corretta prospettiva sulla geometria pseudoeuclidea di Minkowski [32].

Nasce insomma una accezione allargata della geometria spazio-temporale, che si presenta con 4, e non più – per così dire – con 3+1, coordinate. Come lo spazio 3-dimensionale euclideo etichettato mediante il parametro tempo, anche il nuovo spazio 4-dimensionale, o "spazio-tempo" (o "cronotopo" nel linguaggio di certi circoli filosofici rigorosamente spettatori) è privo di curvatura, o piatto, nel senso che in esso esistono sistemi di coordinate per le quali i coefficienti della metrica associata sono costanti; ma a differenza di quanto avviene in quello, tale metrica risulta ora

---

indefinita, con direzioni "spaziali" (per le quali la forma è positiva [33] ), direzioni "temporali" (forma negativa), e direzioni "luminali" o "isotrope" (forma nulla). La geometria di questo spazio non euclideo (propriamente, "pseudoeuclideo di dimensione 4 ed indice 3" con la convenzione qui adottata [34] ) presenta aspetti fortemente eterodossi rispetto alla geometria euclidea: ad esempio, può succedere che la classica disuguaglianza triangolare euclidea (≡ "la somma delle lunghezze di due lati di un triangolo, possibilmente degenere, è maggiore di, o uguale a, quella dell'altro lato") valga alla rovescia (cioè con "minore" al posto di "maggiore"), ma con la nozione di lunghezza *sostanzialmente riformulata*.

3)     LA RELATIVITÀ GENERALE

Lo spazio minkowskiano, "piatto" nello stesso senso in cui lo è quello euclideo, si proponeva come modello di uno spazio-tempo fisico *illimitatamente vuoto di massa-energia, impulso* (≡ momento) *e loro flussi*. Vale a dire, le masse-energie e gli impulsi della dinamica relativistica speciale, e le intensità dei campi elettromagnetici (che attraverso ben note formule equivalgono appunto a densità di energia e impulso), e quindi le cariche/correnti che ne sono le sorgenti lineari, dovevano pensarsi come abbastanza piccole per non infirmare la validità del modello. Appariva dunque naturale, dopo il 1905, l'obiettivo di superare questa limitazione: un'impresa formidabile cui lo stesso Einstein cominciò a dedicarsi molto presto (sembra già intorno al 1907), e che sarebbe stata da lui portata a sostanziale compimento nel 1915-16 [35]. Due intuizioni complementari andarono a poco a poco delineandosi nel progetto einsteiniano. Da una parte vi era l'idea che le densità di energia e impulso, e i loro flussi, distribuiti nello spazio-tempo, ne alterassero in qualche modo la piattezza, "incurvandolo" nel senso tecnico formulato da Gauss per le superficie e generalizzato da Riemann per le generiche varietà con metrica (non degenere) arbitraria. Dall'altra, Einstein immaginò che questa deformazione influisse sul moto [36] dei corpi materiali e dei campi elettromagnetici localizzati, più o meno come le ondulazioni di una superficie influenzano il moto inerziale di un punto materiale ad essa vincolato senza attrito, cioè soggetto alla sola reazione vincolare normale alla superficie stessa. (Questa è soltanto una nota immagine di immediata suggestione didattica, mentre la realtà, in pratica equivalente ad una completa "geometrizzazione" della dinamica, è assai più difficile da rappresentarsi; non tanto perché si deve passare da due a quattro dimensioni, ma perché una delle dimensioni è il tempo.)

---

[33] Questa scelta è convenzionale, e potrebbe essere rovesciata, come molti preferiscono, usando la forma di segno opposto, per cui sono spaziali le direzioni che rendono *negativa* la forma, ecc.

[34] Per comodità del lettore, ricordiamo che l'indice $\pi$ di una forma quadratica non-degenere in n indeterminate $\xi^i$ (i = 1, .., n) è il numero dei suoi addendi positivi dopo che, mediante una opportuna trasformazione lineare non-singolare delle sue indeterminate, diciamo $\xi^i \rightarrow \Xi^i$, essa assume la forma canonica ("pseudopitagorica") $\sum_i \varepsilon_i (\Xi^i)^2$ (somma su i da 1 a n), nella quale ciascun $\varepsilon_i$ è uguale a +1 o a −1. Quindi $\pi = \sum_i (1+\varepsilon_i)/2$. Similmente, il numero degli addendi negativi è $\nu = \sum_i (1-\varepsilon_i)/2$; e ovviamente $\pi+\nu$ = n. La possibile degenerazione della forma consiste nella violazione della precedente ultima relazione in favore della $\pi+\nu$ < n, fermo restando il significato di $\pi$ e di $\nu$ (numero di addendi positivi e rispettivamente negativi della forma resa pseudopitagorica).

[35] Si veda soprattutto la sintesi: A. Einstein, "Die Grundlagen der allgemeinen Relativitätstheorie", Annalen d. Physik **49**, 769 (1916).

Nel suo prevalentemente solitario lavoro, Einstein si ispirò in modo sostanziale ad un cosiddetto "principio di equivalenza" (tra gravitazione e inerzia) alla cui radice è l'identità, in un conveniente sistema di unità di misura, tra la "massa gravitazionale" (cioè la massa come soggetto-oggetto di gravità) e la "massa inerziale di quiete" (il parametro > 0 che figura nelle equazioni dinamiche). Del principio di equivalenza sono note almeno tre versioni progressivamente più forti, la cui formulazione non è sempre priva di ambiguità (equivalenza "debole" – appena formulata –, "media" e "forte"). Secondo il principio di equivalenza forte, «in ogni punto dello spazio-tempo in (presenza di) un arbitrario campo gravitazionale è possibile scegliere un sistema di coordinate "localmente inerziali" tale che, in una regione (spazio-temporale) abbastanza piccola attorno al punto in oggetto, tutte le leggi naturali prendono la stessa forma come in sistemi di coordinate cartesiane non accelerati e in assenza di gravità.» [37] In particolare, il moto di un punto materiale soggetto ad un dato campo gravitazionale, e ad esso soltanto, almeno nella scala spazio-temporale locale è rettilineo ed uniforme in un sistema di riferimento solidale con un corpo materiale in caduta libera nello stesso campo (il sistema localmente inerziale); come è vero per un punto materiale sottratto all'azione di qualunque forza nei comuni riferimenti inerziali della relatività speciale, o della stessa dinamica newtoniana. A completamento del modello, vi era l'esigenza che fossero soddisfatte le condizioni seguenti: (I) la nuova teoria generale della gravitazione si deve ridurre a quella newtoniana per un punto materiale di lagrangiana (differenza tra energia cinetica e energia potenziale) molto più piccola del prodotto della sua massa classica per il quadrato della velocità della luce; (II) in assenza di (densità di) energia-impulso in *tutto* lo spazio-tempo (considerato come varietà 4-dimensionale lorentziana abbastanza regolare), si deve essere ridotti al quadro previsto dalla relatività speciale; e infine (III) le equazioni della teoria devono esprimere l'annullarsi di certi campi tensoriali in tutto lo spazio-tempo, e quindi rimangono formalmente invariate a fronte di arbitrarie trasformazioni regolari delle coordinate (il cosiddetto principio di "covarianza").

---

[36] Più precisamente, "determinasse il moto a meno delle condizioni accessorie". La stessa osservazione vale poco più sotto, allo stesso proposito.
[37] S. Weinberg, Gravitation and Cosmology, Wiley (1972), p. 68.

Quanto agli aspetti matematici del problema, apparve ad Einstein sempre più chiaro che l'elaborazione formale delle sue intuizioni non poteva prescindere da quel "Calcolo Differenziale Assoluto" (di campi tensoriali su varietà pseudoriemanniane 4-dim di indice 3), oggi comunemente noto come "Analisi Tensoriale Classica", che Ricci-Curbastro aveva cominciato a sviluppare sin dal 1892,[38] sulle tracce degli studi di Gauss, Riemann e Christoffel, e che sarebbe stato definitivamente messo a punto con la collaborazione, e alla fine per opera, del suo ex-allievo Levi-Civita.[39] Caratteristica centrale dell'analisi tensoriale è la sua natura "assoluta", cioè il fatto che le sue equazioni si conservano *formalmente invariate*, nel senso della condizione (III) del precedente paragrafo, al cambiare del sistema di coordinate in cui si opera; per cui diventa superfluo, se non per finalità specifiche, riferirsi ad un sistema di coordinate piuttosto che ad un altro. In questo senso, l'analisi tensoriale costituisce insomma l'estrema glorificazione e al contempo la caduta della coordinatazione, la grande intuizione di Cartesio e di Fermat: gli sviluppi della cosiddetta "Analisi su Varietà Differenziabili (astratte)" della prima metà del secolo XX si pongono infatti al di là di questo pur fondamentale strumento concettuale (Analisi Tensoriale – su varietà – cosiddetta "senza coordinate").

Nel seguito gli indici latini variano su 1, .., 4 e quelli greci su 1, .., 3. Nelle sue applicazioni dell'analisi tensoriale alla geometria differenziale delle ipersuperficie immerse in uno spazio euclideo (o più tardi pseudoeuclideo), Ricci aveva incontrato parecchie loro interessanti proprietà. In particolare, egli aveva trovato che, in forza delle cosiddette "identità di Bianchi" – vincoli differenziali lineari del primo ordine tra le $n^2(n^2-1)/12$ (20 per n = 4) componenti algebricamente indipendenti del 4-tensore di Riemann $\rho_{(4)}$ –, la combinazione lineare 2-tensoriale simmetrica $E_{(2)} =: \rho_{(2)} - g_{(2)}\rho_{(0)}/2$ delle due tracce di $\rho_{(4)}$ (la traccia 2-tensoriale $\rho_{(2)}$ di componenti controvarianti $\rho^{jh} \equiv g_{ik}\rho^{jikh}$, o "2-tensore simmetrico di Ricci", e l'invariante lineare $\rho_{(0)} \equiv g_{jh}\rho^{jh}$ di quest'ultimo) era solenoidale, cioè aveva le sue 4 divergenze identicamente nulle ($g_{(2)}$ essendo al solito il tensore fondamentale della varietà).

Pur mancando di una preparazione e di una mentalità specificamente matematiche, Einstein si rese conto dell'importanza che i risultati di Ricci e Levi-Civita potevano avere ai suoi fini, e si sottopose ad un duro tirocinio per assimilare le tecniche del loro calcolo assoluto[40]. Dopo una serie

---

di infruttuosi tentativi, egli imboccò finalmente la strada giusta per stabilire un collegamento tra la metrica dello spazio-tempo e la distribuzione in esso di energia e impulso. Precisamente, egli finì col fissare la sua attenzione sul soprannominato 2-tensore simmetrico-solenoidale $E_{(2)}$ (detto più tardi "tensore di Einstein", da cui il simbolo qui usato per denotarlo, o anche "tensore gravitazionale"), che in forza delle sue proprietà e di certe altre ragioni appariva un candidato ideale per essere supposto "universalmente" proporzionale al 2-tensore energetico $T_{(2)}$, espressione delle densità di energia-impulso e dei loro flussi nello spazio-tempo, e anch'esso simmetrico-solenoidale per definizione. Questo passo decisivo fu compiuto poco tempo prima della vittoriosa conclusione di una ricerca quasi decennale, conclusione che fu resa pubblica con una storica comunicazione all'Accademia Prussiana delle Scienze di Berlino il 25 novembre 1915 .

A partire dalla fine del 1914, e soprattutto nell'ultimo periodo prima del successo, assai fecondo per Einstein fu l'intenso scambio di idee avuto con i due matematici che con più interesse seguivano i suoi progressi nell'"Entwurf" (come Einstein stesso lo chiamava) della formulazione delle equazioni del campo gravitazionale; e cioè Levi-Civita (da Padova) e soprattutto Hilbert (da Gottinga). In particolare l'interazione Einstein-Hilbert è di tale importanza per la storia, e secondo alcuni per lo stesso positivo coronamento, della teoria della relatività generale, che sembra opportuno darne conto in modo abbastanza dettagliato.

A cavallo tra giugno e luglio del '15, e su invito di Hilbert, Einstein trascorse una settimana a Gottinga, dove secondo il suo stesso ricordo tenne un ciclo di «sei seminari di due ore ciascuno» sulla relatività generale, soffermandosi in particolare sull'ancora aperto problema delle equazioni del campo. [41] Se da una parte questo incontro coinvolse definitivamente Hilbert nella ricerca einsteniana, dall'altra esso è all'origine di uno dei più intriganti "gialli" della storia della Scienza. Da quel momento, Hilbert si impegnò infatti in una sua personale ricerca sul problema delle sopraddette equazioni, del quale giunse a soluzione verso il 12/13 novembre, dandone immediata notizia in un pubblico seminario a Gottinga il successivo 16 (seminario al quale Einstein *non fu* presente, nonostante il caloroso invito ricevuto). Il relativo articolo venne sottoposto per pubblicazione il 20 novembre, e apparirà a stampa nel marzo successivo con il titolo un po' enfatico "Die Grundlagen der Physik". [42] Ma intanto, il 18 novembre, in un seminario all'Accademia Prussiana delle Scienze, Einstein comunicava la giustificazione della precessione del perielio di

---

sviluppato un così grande rispetto per la matematica, le cui parti più sottili, nella mia ignoranza, avevo in passato considerato come un semplice lusso! In confronto a questo problema, la vecchia teoria della relatività speciale è un gioco da bambini.» (da una lettera a Sommerfeld del 1912).

[41] Sarebbe di non piccolo interesse, per la storia della Scienza, disporre di una registrazione di quelle 12 ore di lezione einsteiniana. Non abbiamo idea di quanto la sua udienza capì; ma almeno Hilbert si rese lucidamente conto dell'immensa importanza del problema che il grande fisico andava laboriosamente illustrando. Come vedremo, di fatto Hilbert capì "anche troppo".

[42] D. Hilbert, Nachr. Ges. Wiss. Göttingen **3**, 395 (1916); vedi anche lo stesso giornale **1**, 53 (1917), nonché in Math. Annalen **92**, 1 (1924).

Mercurio sulla base di una versione ancora provvisoria delle *sue* equazioni del campo; e pochi giorni dopo, appunto il 25 novembre, si ripresentava all'Accademia con la versione corretta e definitiva delle stesse equazioni. [43] Va da sé che le equazioni di Hilbert e quelle di Einstein risultano equivalenti, pur avendo una differente genesi. Paradossalmente, né Hilbert né Einstein conoscevano le identità di Bianchi, e quindi sulle prime essi non si resero conto che $E_{(2)}$ era *automaticamente* solenoidale; ma mentre il primo *scoprì* da sé questa proprietà, chiudendo finalmente il cerchio, il secondo continuò a credere per qualche tempo che essa dovesse *imporsi* a $E_{(2)}$ per soddisfare alle equazioni del campo (in quanto $T_{(2)}$ era assunto solenoidale).

Vi è insomma abbastanza, nei fatti e nei documenti menzionati, per sollevare una questione non trascurabile (almeno agli occhi degli storici della Scienza): *chi è il vero padre della teoria della relatività generale?* La querelle si è riaccesa in tempi recenti con toni vivacemente polemici. [44] Il nodo della discussione risiede, e probabilmente rimarrà, nella imperfetta conoscenza che abbiamo di quanto i progressi di Hilbert *influenzarono realmente, momento per momento*, quelli di Einstein, e viceversa. Sia come sia, una attenta ricognizione dei documenti originali mostra che i due giunsero in pratica simultaneamente agli stessi risultati sostanziali, pur seguendo linee di pensiero diverse: di impronta decisamente assiomatico-deduttiva quella di Hilbert, più fisica e induttiva quella di Einstein. Un fatto è inoltre certo (lettere di Einstein agli amici H. Zangger e P. Ehrenfest), e cioè che mentre Hilbert sembrava capire benissimo il lavoro di Einstein, quest'ultimo aveva delle difficoltà con quello di Hilbert, trovandolo «oscuro e troppo audace». Aggiungiamo ancora una considerazione, e cioè che se Hilbert si fosse posto per suo conto il problema delle equazioni del campo gravitazionale nei termini in cui glielo pose Einstein in quella fatale estate del '15, egli avrebbe certamente scoperto la relatività generale in modo autonomo, forte delle sue eccezionali capacità matematiche e della sua padronanza del calcolo delle variazioni e dell'algebra degli invarianti tensoriali (per inciso, a Gottinga Hilbert era anche in stretto contatto con E. Nöther, grande esperta dell'uno e dell'altra). Su questa base, sarebbe forse corretto attribuire al binomio Einstein-Hilbert, piuttosto che al solo Einstein, se non l'intera teoria della relatività generale almeno le equazioni del campo che ne sono il naturale coronamento; ma ben pochi, come appunto gli autori appena citati, lo fanno.[45]

---

[43] A. Einstein, "Feldgleichungen der Gravitation", Preuss. Akad. Wiss. Sitz. (pt 2), 844; l'articolo completo è quello già citato negli Annalen d. Physik **49** del 1916, possibilmente da completare con "Hamiltonsches Prinzip und allgemeine Relativitätstheorie", Preuss. Akad. Wiss. Sitz. **2**, 1111 (1916).

[44] La più aggiornata referenza in tal senso è probabilmente "How were the Hilbert-Einstein Equations discovered?" di A.A. Logunov et al., arXiv:physics/0405075 v3, 2004 (41 pp, con la bibliografia essenziale).

[45] La vicenda provocò anche un certo risentimento da parte di Einstein, che inizialmente si sentì defraudato dal celebre collega di una sua pretesa esclusiva priorità; ma il dissapore fu rapidamente superato con l'alta dose di fair play che ci si poteva attendere dalla statura scientifica e umana dei due grandi uomini (come testimonia tra l'altro la loro corrispondenza privata). Comunque, nonostante i loro ottimi rapporti e la reciproca stima, e come era del resto naturale alla luce delle loro diverse attitudini e formazioni, essi non colmarono mai la distanza che separava i loro criteri di

Dobbiamo adesso essere un po' più specifici, seppure al prezzo di proporre un po' informalmente concetti che potrebbero essere definiti con maggior precisione e rigore. Ricordiamo innanzitutto che secondo il modello relativistico-generale lo spazio-tempo si identifica con una varietà 4-dim pseudoriemanniana di indice 3, $L^4_3$, in generale non elementare, e di metrica $g_{(2)}$ non degenere, cioè con segnatura lorentziana [46] (L sta appunto per Lorentz) convenientemente regolare. Tralasciando qui di occuparci delle presumibili proprietà *globali* di questa varietà, [47] la teoria relativistica generale *locale* (cioè limitata ad un aperto $U \subset L^4_3$ dominio di una carta di coordinate $x = (x^1, .., x^4)$), è sostanzialmente ispirata (oltre che al principio di equivalenza) alle due già menzionate idee complementari: (i) «la geometria intrinseca di U, definita dal (campo del) 2-tensore metrico $g_{(2)}$, è determinata, a meno di condizioni accessorie, dal (campo del) 2-tensore energetico totale $T_{(2)}$, supposto simmetrico e solenoidale»; e (ii) «il moto di un punto materiale test di massa trascurabile [48], tra due punti di U congiungibili mediante traiettorie "materiali" ($\equiv$ aventi tangenti *interne* al cono locale con vertice nel punto considerato), avviene lungo una geodetica tra quei punti».

Per quanto ragionevoli, la simmetria e la solenoidalità di $T_{(2)}$ sono essenzialmente postulate. È tuttavia interessante osservare che il semplice modello di una polvere materiale disgregata distribuita in U con densità di energia $c^2\mu^o$ ($\mu^o \equiv$ densità di massa di quiete) e 4-velocità $u^i = cdx^i/\sqrt{(-ds^2)}$ (ds = $|g_{ik}dx^i dx^k|^{1/2}$ e quindi $u_i u^i = g_{ik}u^i u^k = -c^2$, somma sugli indici ripetuti), suggerisce l'espressione

(1)        $T^{ik} =: \mu^o u^i u^k$

---

[46] approccio ai problemi fisico-matematici. È anche noto che Hilbert soleva dire in tono scherzoso, parafrasando un detto memorabile in cui ci si riferiva a guerre e a generali, che «die Physik ist für die Physiker viel zu schwer». Al di là della battuta, naturalmente egli intendeva con ciò affermare che i modelli elaborati e proposti dai fisici sollevano spesso, *una volta matematizzati*, questioni che superano di molto la loro competenza matematica. Hilbert aveva perfettamente ragione: il problema esiste, ed è oggi più che mai attuale e drammatico. Esso reclamerebbe infatti quella più efficace cooperazione tra le due comunità – dei fisici teorici da una parte e dei matematici dall'altra – che invece di rafforzarsi si fa di giorno in giorno obbiettivamente più difficile. Sulla carta, i fisici-matematici dovrebbero colmare le distanze; ma quasi mai ci riescono fino in fondo, forse perché la maggior parte di loro non è capace di stare veramente "nel mezzo" tra le due squadre (un'impresa in ogni caso molto difficile).

[46] Anche nel caso pseudoriemanniano, la segnatura della metrica si riferisce ai segni dei suoi coefficienti una volta che essa sia ridotta *localmente* a forma canonica mediante trasformazioni delle coordinate (teorema di Sylvester). Con la nostra convenzione, dire che la metrica è lorentziana significa che dei quattro coefficienti diagonali tre sono positivi e uno negativo. Il numero dei coefficienti positivi figura in basso nella notazione $L^4_3$.

[47] Quello delle proprietà globali della varietà spazio-temporale $L^4_3$ diventerà presto il problema fondamentale della cosmologia relativistica-generale (usiamo questa espressione anche se dall'entrata in scena delle teorie quantistiche è diventato impossibile parlare di una cosmologia *interamente* ispirata ad una teoria macroscopica qual è la relatività generale). Va anche ricordato che, come era naturale, l'approccio einsteiniano alla sua teoria fu di tipo locale. L'interesse per i problemi globali su varietà astratte generiche, e soprattutto gli strumenti matematici necessari per affrontarli, avevano cominciato a svilupparsi seriamente poco prima, anche indipendentemente dalla relatività generale.

[48] La massa del punto test deve essere supposta piccola per non perturbare la metrica di riferimento attraverso le equazioni del campo gravitazionale (6) che seguono. Ciò non significa che, venendo meno questa condizione "adiabatica", il movimento del punto materiale non sarebbe ugualmente determinato: il suo calcolo sarebbe tuttavia assai più difficile, dovendosi allora determinare in modo consistente sia la metrica che il movimento del punto.

per il relativo tensore energetico di $T_{(2)}$. [49] Richiedendo allora la solenoidalità di questo tensore simmetrico, cioè le

(2)     $(\mu^o u^i u^k)_{/i} = 0$

(dove $_{/i}$ denota la derivata covariante rispetto a $x^i$), si ottiene

(3)     $(\mu^o u^i)_{/i} = 0$, [50]

e quindi

(4)     $u^i u^k_{/i} = 0$

se $\mu^o \neq 0$. Queste ultime (4) mostrano che il campo di 4-velocità è geodetico, cioè si mantiene tangente a una geodetica di U, e quindi che la generica particella di polvere materiale si muove lungo questa geodetica. In altre parole, nel modello considerato la solenoidalità di $T_{(2)}$ implica la (ii) del paragrafo precedente; e viceversa, la (ii) stessa implica la solenoidalità di $T_{(2)}$ sotto la condizione (3) (che il vettore 4-dim $\mu^o u^i$ sia solenoidale). Questo è conforme al cosiddetto "principio della geodetica spazio-temporale" della relatività generale, che degenera nella legge d'inerzia della relatività speciale (e anche della dinamica newtoniana) per $T_{(2)} = 0$. Quanto alla (3), nel limite classico ($u^{1 \leq i \leq 3} \approx u^i$, $u^i$ essendo le componenti della 3-velocità della generica particella, e $u^4 \approx c$) essa si riduce alla legge di conservazione di massa $\text{div}(\mu^o u) + \partial_t \mu^o = 0$.

Come sappiamo, le $4(4+1)/2 = 10$ componenti algebricamente indipendenti di $T_{(2)}$ rappresentano la densità di energia ($i = k = 4$, una quantità scalare), di impulso o momento ($i = 1,2,3$, $k = 4$, tre quantità scalari) e i loro flussi ($i, k = 1,2,3$, sei quantità scalari). In forza del teorema di Gauss generalizzato, le

(5)     $T^{ik}_{/k} = 0$

assicurano allora che le corrispondenti quantità integrali (cioè l'energia totale, l'impulso totale, ecc., in U, o nella sua parte di interesse) "si conservano". Per questa ragione le (5) sono dette "equazioni di conservazione".

L'anello mancante in questo quadro è quello di rendere concreto ed effettivo il principio (i) (di tre paragrafi più sopra), plausibilmente sotto la forma di un sistema di 10 equazioni differenziali nelle 10 incognite $g_{ik}$. Ora le 20 componenti del tensore di Riemann $\rho_{(4)}$ si esprimono come funzioni di quelle del tensore fondamentale $g_{(2)}$ e delle loro derivate prime e seconde, essendo lineari in queste ultime. Lo stesso vale per le 10 componenti del 2-tensore-traccia $\rho_{(2)}$ di $\rho_{(4)}$, simmetrico ma

---

[49] Vedi il già citato "The meaning of Relativity", in cui, dopo aver calcolato l'espressione del tensore energetico *elettromagnetico* (nell'ambito della relatività ristretta), Einstein ne verifica il carattere simmetrico e solenoidale esternamente alle cariche elettriche (vedi l'eq. (47c) del testo einsteiniano), aggiungendo poi: «È molto difficile evitare di fare l'ipotesi che, anche in tutti gli altri casi, la distribuzione spaziale dell'energia sia data da un tensore simmetrico $T_{\mu\nu}$, e che tale tensore soddisfi la relazione (47c)». Immediatamente dopo, Einstein considera il caso della polvere materiale e dà l'espressione (1) del relativo tensore energetico (sua eq. (50)).

[50] Basta sviluppare il 1° membro della (2) e contrarlo con $u_k$. Questo dà $0 = -(\mu^o u^i)_{/i} + c^{-2}\mu^o u^i u_k u^k_{/i} \equiv -(\mu^o u^i)_{/i}$.

*non* necessariamente solenoidale, nonché per le altrettante componenti del 2-tensore gravitazionale $E_{(2)}$, simmetrico *e* solenoidale. La già accennata idea di Einstein, maturata dopo lunga e strenua ricerca, fu quella di postulare la proporzionalità di $E_{(2)}$ a $T_{(2)}$ secondo un fattore costante universale che diremo qui $-8\pi\kappa/c^4$, con $\dim(\kappa) = M^{-1}L^3T^{-2}$ [51]. Sotto la naturale richiesta che valga un "principio di corrispondenza" per il quale la dinamica einsteiniana deve confluire in quella newtoniana al più basso ordine significativo in $\varphi/c^2$ (con $\varphi \equiv$ energia potenziale gravitazionale newtoniana per massa unitaria) si riconosce allora che la costante $\kappa$ deve identificarsi con la costante gravitazionale di Newton-Cavendish [52]. In conclusione, le equazioni einsteiniane si scrivono nella forma

(6)      $E_{(2)} + (8\pi\kappa/c^4)T_{(2)} = 0$,

dove $E_{(2)}$ è una ben definita funzione di $(g_{(2)},\partial g_{(2)},\partial^2 g_{(2)})$, lineare nelle derivate seconde $\partial^2 g_{(2)}$, e $T_{(2)}$ è una funzione data (via modellazione) di $(x,g_{(2)},\partial g_{(2)})$. [53] Le 10 equazioni (6) nelle altrettante incognite $g_{(2)}$ costituiscono un sistema differenzial-parziale quasi-lineare del 2° ordine, che si riconosce di tipo iperbolico. Il bilancio tra equazioni ed incognite nelle (6) è corretto ad un esame superficiale (10 contro 10); ma tale rimane anche tenendo conto della solenoidalità dei due membri delle (6) (6 contro 6). Infine si dovrà presupporre che la soluzione $g_{(2)}$ delle (6) e relative condizioni accessorie rispetti la natura postulata di campo 2-tensoriale *uniformemente* non degenere (o regolare) e lorentziano (la prima condizione assicura la seconda se questa vale in un punto della varietà, assunta connessa).

Naturalmente ci si aspetta, e così risulta essere, che la funzione $E_{(2)}$ di $g_{(2)}$ e delle sue derivate prime e seconde si annulli se queste derivate prime e seconde si annullano. Questo fatto assicura la necessaria coerenza con il quadro della relatività speciale, in cui per definizione il tensore energetico $T_{(2)}$ è nullo in *tutto* lo spazio-tempo; ma non è detto che valga il contrario in un dominio *limitato* e semplicemente connesso, cioè che "$T_{(2)} \equiv 0$" in quel dominio implichi che esista un riferimento per il quale $g_{(2)}$ sia ivi costante. (La precisa risposta a ciò si trae dallo studio del corrispondente problema di Cauchy per il sistema (6).) Si osservi ancora che il sistema (6) è non-lineare non soltanto perché le $g_{(2)}$ e le $\partial g_{(2)}/\partial x$ compaiono in modo non-lineare in $E_{(2)}$, ma anche, possibilmente, in $T_{(2)}$; e infine, che la possibile dipendenza di $T_{(2)}$ da $g_{(2)}$ e derivate prime fa sì che il tensore energetico non possa, in generale, trattarsi come "termine libero" del sistema. [54]

---

[51] Le componenti (di qualunque tipo) del tensore energetico hanno le dimensioni di una densità di energia (cioè $ML^{-1}T^{-2}$) se quelle del tensore metrico sono, come è norma supporre e supporremo, adimensionali. Anche le componenti $dx^i/ds = u^i/c$ risultano allora adimensionali. Su questa base, si verifica facilmente la correttezza della $\dim(\kappa)$ riportata nel testo. (Più in generale, le componenti del tensore energetico hanno le dimensioni del prodotto di una densità di energia per le componenti omologhe – cioè della stessa natura covariante, controvariante o mista – del tensore metrico, o se si preferisce delle componenti omologhe del divettore delle 4-velocità.)

[52] Un valore abbastanza aggiornato di questa costante è di $6,6726 \cdot 10^{-11}$ m³kg⁻¹s⁻² (G. Luther, W. Towler, 1982). Oltre che "di Newton", la costante gravitazionale si dice anche "di Cavendish" dal nome del fisico inglese (1731-1810) che per primo la misurò nel 1798 trovando il valore incredibilmente preciso di $6,754 \cdot 10^{-11}$ m³kg⁻¹s⁻².

[53] Non ci diffondiamo qui sulla travagliata storia della "costante cosmologica" – riesumata e in gran parte riabilitata in tempi più recenti –, che fu presto aggiunta da Einstein nelle (6) nell'intento di giustificare un modello *stazionario* dell'universo.

La soluzione del sistema differenzial-parziale (SDP) iperbolico quasi-lineare del 2° ordine (6) sotto le convenienti condizioni accessorie costituisce il problema per eccellenza della teoria della relatività generale, oggetto ancora oggi di attiva ricerca analitica e computazionale. Un caso di grande interesse è in particolare quello del SDP ovunque (nel dominio convenuto) omogeneo associato al SDP (6), le cui soluzioni dipendono quindi esclusivamente dalle condizioni accessorie (iniziali e al contorno), ed hanno tipicamente natura di propagazione ondosa. È molto significativo che il modulo della velocità di avanzamento normale in un punto del fronte di queste onde di gravità, ipersuperficie ($\equiv$ superficie mobili) caratteristiche del SDP (6) attraverso le quali le derivate seconde di $g_{(2)}$ possono essere discontinue (restando $g_{(2)}$ continuo con le sue derivate prime), risulti uguale a quello della velocità della luce in quel punto.

Del SDP (6) si conoscono ormai molte soluzioni esatte, legate ad ipotesi limitative particolari: le più famose, e prime in ordine di tempo, sono le soluzioni statiche ottenute da Schwarzschild (Karl, 1873-1916) nel 1916, nel caso di un punto materiale, e rispettivamente di una palla materiale omogenea incompressibile, l'uno e l'altra immersi nello spazio-tempo vuoto e infinito [55]. Entrambe prevedono la possibile esistenza di corpi non irraggianti alcuna forma di massa-energia – o "buchi neri" (S. Hawking, R. Penrose, 1965-70) – congetturalmente prodotti dal collasso gravitazionale di stelle di massa sufficientemente elevata. Un primo esempio di buco nero fu osservato nel 1971. Altri contributi fondamentali allo studio del SDP (6) furono quelli di Friedmann (Alexandr, 1888-1925), che per primo elaborò un modello cosmologico fondato sulle equazioni einsteiniane (1922, 1924) nel quale la densità di massa media nell'universo era assunta costante e tutti gli altri parametri erano noti salvo la sua curvatura. [56] Il modello di Friedmann implica una dinamica dell'universo del tipo cosiddetto "big-bang" (locuzione dovuta a Hoyle (Fred, 1915-2001)), proposta esplicitamente da Lemaître (Georges, 1894-1966) [57] nel 1927 e ulteriormente studiata da Gamow (Georgy, 1904-1968) intorno al 1948. Esso fu dunque sostanzialmente formulato con circa sette anni di anticipo sulle osservazioni (divenute abbastanza convincenti intorno al 1929, ma iniziate parecchi anni prima) di Hubble (Edwin,1888-1953), che dall'osservatorio di M.te Wilson fornì le prime evidenze sperimentali dell'espansione cosmologica, governata dalla legge di approssimativa proporzionalità tra la velocità di allontanamento – e quindi tra lo spostamento verso il rosso della luce da loro emessa – e la distanza delle galassie lontane, nota appunto come legge di Hubble. [58]

---

[54] Si dimostra che le equazioni associate omogenee ($T_{(2)} \equiv 0$) delle (6), nel dominio U, equivalgono al seguente principio variazionale: «L'integrale $\int_U \rho_{(0)} |g|^{1/2} dx$ è stazionario quando si richieda che le variazioni di $g_{(2)}$ e delle sue derivate prime standard (quindi dei coefficienti di Christoffel di 1ª specie $\Gamma_{ikj}$ ) si annullino sul contorno $\partial U$ di U»; dove $\rho_{(0)}$, ricordiamo, è la traccia scalare del tensore di Riemann, e $|g|$ sta per det$\{g_{ik}\}$. Questo principio, una specializzazione del principio di azione stazionaria che Hilbert usò nella sua deduzione delle equazioni del campo, avrà un significativo ruolo ispiratore nei successivi e infruttuosi tentativi di Einstein (e non solo di lui) di costruire una "teoria unitaria", cioè capace non solo di *rappresentare* ma anche di *geometrizzare* i fenomeni elettromagnetici, come la relatività generale fa con quelli gravitazionali, in uno spazio-tempo *non pseudoriemanniano*.
[55] K. Schwarzschild, "Über das Gravitationsfeld eines Massenpunktes nach der Einsteinschen Theorie", Sitz. Preuss. Akad. Wiss. 189 (1916); "Über das Gravitationsfeld einer Kugel aus inkompressibler Flüssigkeit nach der Einsteinschen Theorie", Sitz. Preuss. Akad. Wiss. 424 (1916).
[56] A. Friedmann, "Über die Krümmung des Raumes" Z. Physik **10**, 377 (1922); "Über die Möglichkeit einer Welt mit konstanter negativer Krümmung des Raumes", Z. Physik **21**, 326 (1924).

Vari ardui problemi della cosmologia contemporanea restano aperti: ad esempio il problema della densità media di materia/radiazione, che discrimina tra un universo "chiuso" (che nasce da una esplosione e termina con una implosione), ed un universo "aperto" in indefinita espansione (possibilità previste da Friedmann, e legate alla curvatura del cosmo), la questione della costante cosmologica che sebbene "appena" diversa da zero interpreta l'esistenza di una "energia del vuoto", .. e via elencando. Almeno in parte, essi sono legati alla mancanza di una teoria unitaria nella quale coabitino la fisica quantistica (microscopica) e la relatività generale (macroscopica) [59], e quindi alla nostra conoscenza indiretta e congetturale, accettando, come sembra da tempo necessario, il modello singolare del big-bang della fisica dei primi istanti di vita dell'universo. Inoltre non è detto che le condizioni iniziali al momento del big-bang abbiano necessariamente lasciato tracce significative nell'universo attuale: un breve periodo di espansione rapidissima, o "inflazione" (A. Starobinskij, A. Guth), potrebbe averle quasi del tutto cancellate. (Di fatto, l'ipotesi inflativa ha ormai una solida posizione all'interno dell'odierno "modello cosmologico standard"). Ma allora diventa naturale chiedersi quale microfisica rese possibile o determinò l'inflazione. Il problema delle condizioni iniziali torna così in primo piano, seppure in una accezione ben diversa da quella che si sarebbe formulata in ambito relativistico-generale puro. [60] Sottolineiamo qui ancora (cfr. nota ([47])) che un ipotetico tentativo di illustrare la cosmologia "moderna" – cioè, sviluppata all'incirca dopo il terzo decennio del '900 – in termini strettamente macroscopici sarebbe un non-senso logico e storico.

---

[57] G. Lemaître, Ann. Soc. Sci. Bruxel., **A47**, 49 (1927) ; Mon. Not. Roy. Astron. Soc., **91**, 483 (1931). Lemaître parlò di «atom primitif» a proposito di quello che si sarebbe poi detto "big-bang", ma la sua espressione non ebbe successo, sebbene non fosse impropria. L'idea di chiamare (sprezzantemente) big-bang la grande esplosione iniziale, ormai definitivamente affermata, risale a F. Hoyle, l'uomo che più tenacemente avversò il modello singolare dell'universo.

[58] Se ci si riferisce alla notissima immagine intuitiva del palloncino di gomma che si gonfia progressivamente (ecco uno splendido esempio di didattica metaforica!) si intuisce che dei segni praticati sulla sua superficie si allontanano l'uno dall'altro con velocità *crescente* con la loro distanza.

[59] L'analoga unificazione tra fisica quantistica e modello relativistico *speciale* è virtualmente compiuta nella moderna elettrodinamica quantistica. Si potrebbe aggiungere che in ultima analisi la relatività generale è una particolare teoria di gauge (non-abeliana), e quindi sotto questo aspetto è legata ai moderni sviluppi della fisica teorica. La speranza della fisica contemporanea è quella di giungere a descrivere tutte le interazioni fondamentali nell'ambito di una *singola* teoria di gauge unificata.

[60] Tra le altre, una generalizzazione relativistico-generale dell'equazione di Schrödinger è stata proposta J.A. Wheeler e B. De Witt. Ma ovviamente non si ha nessuna conferma osservativa della sua validità.

Altre soluzioni delle equazioni einsteiniane sono quelle di tipo asintotico basate su procedure di linearizzazione ricorsiva, o di linearizzazione pura e semplice, del SDP (6); nonché quelle che possono calcolarsi numericamente in casi particolari. La ricerca fisico-matematica sul sistema delle equazioni del campo gravitazionale (esatte o approssimate) continua ad essere quanto mai viva: si stima che delle *parecchie migliaia* di articoli pubblicati sul solo tema delle sue soluzioni esatte, circa due terzi siano apparsi negli ultimi venti anni. [61]

4)    LA RELATIVITÀ E I FATTI OSSERVATIVI

Ci proponiamo adesso di presentare una succinta rassegna dei fatti sperimentali che furono a fondamento della teoria della relatività (speciale e generale) o piuttosto che ne diedero successive convalide a posteriori. Va infatti subito riconosciuto (o in parte ricordato, alla luce di quanto già esposto) che la relatività nel suo insieme *è un esempio supremo di teoria fisico-matematica fondata su pochissimi fatti empirici specifici,* il cui assetto finale fu essenzialmente ottenuto con il solo mezzo di audaci generalizzazioni e di una possente capacità di sintesi.

Cominciamo con la relatività speciale. Anche se non è chiaro in quale misura, è probabile che lo storico esperimento di Michelson e Morley, nelle sue successive versioni fino a quella del 1887, nonché gli esperimenti alternativi intesi allo stesso scopo e già menzionati, abbiano influenzato l'elaborazione della relatività speciale *meno* di quanto fece, nei suoi creatori [62] , la precognizione della teoria elettromagnetica di Maxwell e del suo *non* essere invariante rispetto alle trasformazioni galileiane delle coordinate e a quelle prerelativistiche dei campi. [63] Come sappiamo, la costanza (del modulo) della velocità della luce nel vuoto, *fondamento empirico centrale* della teoria, è soltanto uno degli ingredienti logici specifici sulla cui base si possono *dedurre* le trasformazioni di Lorentz; esso, e gli altri tre principi (di relatività, di linearità e di isotropia perpendicolare), furono tuttavia essenzialmente *presupposti* da Einstein, come avrebbe potuto esser fatto da chiunque altro sulla base delle stesse conoscenze sperimentali, a tutti accessibili. In altre parole, ciò che distinse Einstein dai fisici a lui contemporanei (con l'eccezione di Lorentz, che lo precedette sotto non pochi fondamentali profili) è *l'uso* che egli fece di quelle conoscenze, nulla di più e nulla di meno.

---

[61] Vedi in particolare la monumentale monografia di H. Stephani, D. Kramer, M. MacCallum, C. Hoenselaers, E. Herlt: "Exact Solutions of Einstein's Field Equations", Cambridge Un. Pr., aggiornata nel 2003.

[62] Vale a dire, innanzitutto Einstein e Lorentz; e secondariamente, sotto il profilo dell'inquadramento matematico, Minkowski. Quanto a Poincaré, sembra chiaro che il fisico-matematico francese non ne capì abbastanza la *base fisica*, restando fedele all'idea dell'elettrone "appiattito" dal vento d'etere. In una lezione tenuta nel 1910, Poincaré si riferì alla scelta tra l'elettrodinamica di Lorentz e la relatività speciale come ad «una questione di gusti».

[63] M. Polany, un noto fisico-chimico che aveva una lunga familiarità con Einstein, nel 1954 pubblicò con l'apparente approvazione di quest'ultimo la seguente discutibile affermazione: «L'esperimento di Michelson e Morley ebbe un effetto trascurabile sulla scoperta della relatività.» E proseguendo, un po' sopra le righe: «La consueta spiegazione data dai manuali di relatività come di una risposta teorica all'esperimento di Michelson e Morley è un'invenzione. È il prodotto di un pregiudizio filosofico. Quando Einstein scoprì la razionalità nella natura, senza l'aiuto di alcuna osservazione che non fosse già disponibile da almeno cinquant'anni, i nostri manuali positivistici coprirono prontamente lo scandalo con una spiegazione convenientemente manipolata della sua scoperta.»

Le convalide sperimentali della relatività speciale sono innumerevoli e, come si dice, ormai "al disopra di ogni sospetto": qualunque trattato istituzionale di fisica non troppo vecchio ne descrive gli esempi più significativi. Tra questi, basterà qui ricordare che la progettazione e realizzazione delle macchine acceleratrici, dagli anni '30 in qua, sarebbe impensabile senza tener conto della dinamica einsteiniana speciale. La ben nota "equivalenza" (≡ uguaglianza a meno del fattore $c^2$) tra massa e energia è invece soltanto *compatibile* con la teoria, perché quest'ultima si limita ad imporne la validità a meno di una costante additiva arbitraria. Quindi in quella equivalenza va visto un addizionale colpo d'ala di Einstein, che la propose in una breve nota interrogativa dello stesso 1905 [64]. La popolarissima [65] uguaglianza "energia e = (massa di moto m) × (quadrato della velocità della luce $c^2$)" che ne traduce il significato, avrà le sue prime conferme (sostanzialmente nella versione "di quiete" $e^o = m^o c^2$) con gli studi sulla radioattività naturale e artificiale, nelle prime precise valutazioni del difetto di massa di certe reazioni nucleari (Cockcroft e Walton, 1932 [66]), e ancora nella interpretazione dei meccanismi che producono lo splendore del sole (C.F Von Weizsäcker e H.A. Bethe, 1938-39 [67]). Altre conferme verranno più tardi dalle reazioni nucleari di fissione (Chicago, 2 dicembre 1942; Alamogordo (USA), 16 luglio 1945), e ancora dopo, di fusione (atollo di Bikini, con oltre venti esplosioni sperimentali tra il 1946 e il 1958).

Ben poco fondato sull'esperimento appare lo sviluppo della relatività generale, fino alla sua sostanziale messa a punto nel 1915/16. Con questa affermazione intendiamo ancora sottolineare che la teoria einsteiniana della gravitazione risultò dalla riflessione profonda, guidata dallo straordinario intuito di una sola mente (o al più di due, volendo mettere nel conto l'importante ma non decisivo contributo di Hilbert), su un ristretto numero di principi fisici ragionevoli. Di questi principi, quello dell'"equivalenza" (debole e forte), quello della necessità che la teoria generale si riducesse alla teoria speciale per tensore energetico totale nullo (e quindi che sia l'esistenza di questo campo tensoriale a provocare il passaggio da uno spaziotempo pseudoeuclideo (piatto) ad uno spaziotempo pseudoriemanniano (curvo)), e infine quello che la teoria gravitazionale relativistica confluisse nella corrispondente teoria newtoniana al 1° ordine per piccolo $\varphi/c^2$, furono probabilmente i più fecondi.

---

[64] A. Einstein, "Ist die Trägheit eines Körpers von seinem Energiegehalt abhängig?", Annal. der Physik **18**, 639 (1905).

[65] Benché la e = $mc^2$ sia diventata ormai una vera e propria icona mediatica, del suo significato è lecito dubitare che sia veramente compreso dalla maggior parte delle persone cosiddette "di media cultura". Come abbiamo già riferito, lo stesso Einstein era ancora perplesso sulla sua interpretazione a molti anni di distanza dalla prima (dubitativa) proposta.

[66] J.D. Cockcroft, G.T. Walton, Proc. Roy. Soc. A, **137**, 229 (1932). Per quanto meno clamorosa delle successive, fu questa la prima vera conferma della e = $mc^2$. Lo storico esperimento consistette nello studio del bilancio di massa della reazione $_3^7Li + _1^1H = 2 _2^4He$, in cui un nucleo di litio-7 reagisce con un protone producendo due particelle α. Nelle unità standard in cui un atomo di ossigeno ha massa 16, la massa di $_3^7Li$ ammonta a 7,0166, quella di $_1^1H$ a 1,0076, e quella di $_2^4He$ a 4,0028. Quindi dopo la reazione manca una massa pari a 0,0186, che deve apparire come energia cinetica delle due α. Tradotta in Joule, questa energia ammonta a 2,781 $10^{-12}$ J. La misura fu condotta col "metodo del range", e confermò in pieno questa previsione. (Nuove misure più precise mettono anche in conto l'energia cinetica del protone incidente.)

[67] La reazione nucleare in oggetto è quella della fusione di quattro $_1^1H$ in un $_2^4He$. In realtà la reazione diretta non è possibile, ma Bethe (che ebbe anche per questo il Nobel nel 1967) suggerì che lo diventasse per il tramite di un nucleo di carbonio-12. Dapprima questo si fonde con tre protoni, producendo un nucleo di azoto-15; successivamente quest'ultimo si fonde a sua volta con il quarto protone, riproducendo l'originale carbonio-12 più il nucleo di elio-4.

A differenza dalla teoria gravitazionale di Newton, che genialmente riassunse i risultati di una lunga ed intensa campagna di osservazioni astronomiche (principalmente ad opera di T. Brahe, J. Kepler e dello stesso Newton); e ancor più, a differenza dalla teoria elettromagnetica di Maxwell (nella quale Cavendish, Coulomb, Ampère, Faraday, e naturalmente lo stesso Maxwell giocarono analogo ruolo di sperimentatori), la relatività generale resta dunque un esempio praticamente unico di teoria fisica "quasi-autoreferenziale" nel preciso senso sopra specificato. Per quanto riguarda il principio dell'equivalenza debole, vi era in realtà la vasta campagna di misure di Eötvös (Roland, 1848-1919), che a cavallo tra i due secoli ne dimostrava la validità con sempre maggior precisione relativa (fino a $10^{-7} \div 10^{-8}$) [68] ; ma è del tutto fuori luogo proporre una possibile connessione genetica da quelle (o altre) misure allo sviluppo della relatività generale. [69] In modo non dissimile da quanto era già occorso con la costanza di c nei confronti della relatività speciale, Einstein partì infatti dal *presupposto*, di per sé abbastanza naturale, della equivalenza debole; della quale si era del resto ben consapevoli anche prima di Eötvös, come la stessa coppia newtoniana {legge dinamica, legge di gravitazione universale} implicava, seppur con una precisione che a quei tempi non superava $10^{-5}$. [70] Insomma, e in maggior misura che nel caso della relatività speciale, lo sviluppo della relatività generale procedette piuttosto da una sofisticata elaborazione concettuale di certe verità sperimentali all'epoca alquanto evidenti che da nuove osservazioni. Alla luce di questa circostanza, è certamente più appropriato occuparsi dei numerosi esperimenti ai quali a ragione si attribuisce un valore di convalida della teoria, che di presunti esperimenti che ne avrebbero in qualche modo "pilotato" la messa a punto a partire dai primi tentativi in tal senso (ca. 1907). [71]

---

[68] R. Eötvös, Math. natur. Berichte Ungarn, **8**, 65 (1891); Ann. Phys. Chem., **59**, 354 (1896).

[69] A proposito del principio di equivalenza debole, lo stesso Einstein ebbe così ad esprimersi (in "On the Origins of the General Theory of Relativity"): «I did not seriously doubt its strict validity even without knowing the result of the beatiful experiment of Eötvös, which – if I remember correctly – I only heard of later».

[70] Campagne sperimentali più recenti e sofisticate hanno migliorato il limite di Eötvös fino a circa $10^{-10}$ (Dicke et al., 1960), e addirittura a $10^{-11} \div 10^{-12}$ (Braginsky e Panov, 1971).

[71] Indubbiamente la relatività generale fornì ampio supporto all'importanza (se non al primato) del momento raziocinante – induttivo e deduttivo – dell'attività scientifica "esatta", rispetto al suo momento strettamente osservativo-sperimentale. Questa opinione era del resto abbastanza in sintonia con lo spirito dei tempi, e fortemente rappresentata dai più autorevoli scienziati-epistemologi a cavallo dei due secoli; in particolare dallo stesso Einstein e da Poincaré. Il quale ultimo ammoniva: «un insieme di fatti è una scienza non più di quanto un mucchio di mattoni sia una casa».

Innanzitutto a se stesso, Einstein suggerì tre diverse "verifiche cruciali" della relatività generale, e cioè:

(a) la precessione del perielio delle orbite dei pianeti, e specialmente di quella di Mercurio, che per le sue caratteristiche meglio si prestava alla possibile ricognizione del fenomeno;

(b) la deflessione dei raggi luminosi provenienti da stelle angolarmente prossime al bordo del Sole (in una data posizione della Terra lungo la sua orbita), e quindi l'apparente spostamento angolare di quelle stelle relativamente ad una situazione di non-allineamento (di fatto una *divergenza* angolare, perché la deflessione deve avere il Sole dalla parte della concavità del raggio luminoso), doppia di quella prevedibile sulla base della sola equivalenza massa-energia della relatività speciale;

(c) lo spostamento verso il rosso (o "redshift") dello spettro di radiazione elettromagnetica emessa da sorgenti massive (come il Sole, o assai meglio, come una "nana bianca", molto più luminosa e pesante), dovuto al loro campo gravitazionale. #

Aggiungiamo qualche commento su questi esperimenti di convalida.

Su (a): Nel corso del XIX secolo, si era accertato [72] che il calcolo delle perturbazioni interplanetarie non rendeva completamente conto della precessione secolare dell'orbita di Mercurio: restava un difetto di circa 43″ d'arco su un totale effettivo di 575″. Già nel 1915, secondo quanto aveva annunciato all'Accademia prussiana delle Scienze, Einstein aveva ricavato dalle sue equazioni, seppur ancora scorrette, una semplice (e corretta) formula approssimata [73] che gli permetteva il calcolo del contributo relativistico-generale alla precessione dell'orbita di un pianeta generico in funzione del suo semiasse maggiore, della sua eccentricità e del suo periodo. Applicata all'orbita di Mercurio, la formula in oggetto dava un'ottima giustificazione della discrepanza osservata.

Su (b): La divergenza angolare apparente – un effetto che può permettere la visione di oggetti luminosi situati marginalmente "dietro" il Sole – fu effettivamente osservata (isola Principe, 29 maggio 1919) in occasione di una eclisse solare totale, e stimata intorno al previsto ammontare di 1,75″ d'arco, seppur con non poche incertezze. (Il mascheramento dell'astro è una condizione praticamente necessaria per osservare oggetti celesti nella direzione della sua altrimenti luminosissima periferia.) Al giorno d'oggi, il sopracitato valore di 1,75″ è sufficientemente confermato da numerose e più precise misure. [74]

---

[72] Protagonista di queste laboriose valutazioni fu quel Le Verrier (Urban, 1811-1877) dell'Osservatorio astronomico di Parigi che di lì a pochi anni sarebbe stato uno degli scopritori del pianeta Nettuno. Nel 1859 egli pubblicò una versione revisionata della teoria della precessione dell'orbita di Mercurio, scoprendovi una discrepanza apparentemente ineliminabile rispetto alle osservazioni del tempo. Quell'anomalia permaneva sostanzialmente immutata (dai 38″ di Le Verrier a quasi 43″) oltre mezzo secolo dopo.

[73] A. Einstein, "Erklärung der Perihelbewegung des Mercur aus der allgemeinen Relativitätstheorie", Preuss. Akad. Wiss. Sitz., pt. 2, 831 (1915).

Su (c): Il redshift della nana bianca 40 Eridani B fu osservato nel 1954 [75], in buon accordo con il valore di circa $10^{-5}$ previsto dalla teoria. Misure di questo tipo sono comprensibilmente delicate, dovendo esse venir depurate dal redshift Doppler dovuto all'espansione cosmologica (altra conferma indiretta della teoria generale, via soluzioni di Friedmann). La disponibilità di una diagnostica estremamente sensibile ha poi permesso di sfruttare il redshift gravitazionale in un brillante esperimento [76] il cui risultato costituisce oggi una delle convalide più attendibili della relatività generale. In esso, vengono confrontate le frequenze di due sorgenti di radiazione elettromagnetica identiche, ma poste a quote diverse, una sulla cima e l'altra alla base di una torre alta circa venti metri. La differenza relativa risulta dell'ordine di $10^{-15}$, ma la sua precisa misura si scosta di un mero 1% da quanto prevede la teoria. #

Altri effetti relativistico-generali furono previsti e verificati con buona o accettabile precisione. Ci limiteremo qui a ricordarne tre, e cioè:

(d) l'esistenza di una "radiazione (elettromagnetica) cosmica di fondo" (CMB, Cosmic Microwave Background)" a ca. 3 K, residuo fossile della congetturale grande esplosione iniziale. La CMB è caratterizzata da un alto grado di isotropia e da una distribuzione energetica tipica dell'equilibrio termico, entrambe previste dai modelli evolutivi del cosmo includenti una fase inflativa;

(e) il ritardo di segnali elettromagnetici dovuto al loro attraversamento di una regione spaziale con forte campo gravitazionale (rispetto alle previsioni che non tengano conto di questo fatto);

(f) la possibile formazione di immagini multiple di oggetti celesti situati dietro grandi galassie, e più in generale, la possibilità che queste possano comportarsi come vere e proprie "lenti" attraverso le quali osserviamo oggetti più lontani. #

Come già (a) e (b), (e) e (f) devono classificarsi come effetti "ottici" della gravità. Seguono brevi commenti su (d), (e) e (f).

---

[74] Vedi G.C. Mc Wittie, "General Relativity and Cosmology", Wiley (1956), per le analoghe osservazioni effettuate in concomitanza con eclissi solari dal 1919 al 1952. Due altre osservazioni dello stesso tipo erano state tentate *prima* del 1919, nel 1912 in Brasile e nel 1914 in Crimea; ma entrambe si conclusero con un nulla di fatto, a causa delle avverse condizioni atmosferiche la prima, e dello scoppio della guerra mondiale la seconda. Il fallimento di questi due primi esperimenti sulla deflessione della luce ad opera della gravità solare fu in fondo un bene per Einstein, perché a quei tempi egli non aveva ancora sviluppato completamente la teoria relativistica della gravitazione, e la sua previsione sull'ammontare della deflessione la sottostimava di un fattore 2. Tra le convalide della relatività generale, la "retrodizione" della precessione di Mercurio (fu infatti fondata sulle effemeridi astronomiche già disponibili piuttosto che su nuove osservazioni), oltre che prima in ordine di tempo, fu anche molto più solida della "predizione" della deflessione della luce da parte del sole; ma curiosamente (anche se non troppo), mentre la notizia della prima rimase confinata agli ambienti scientifici, la seconda fu investita da un enorme clamore mediatico, che fece di Einstein una vera e propria star internazionale.

[75] D.M. Popper et al., Astr. Journ. **120**, 316 (1954). Per simili studi su nane bianche, vedi J.L. Greenstein, V. Trimble, Astr. Journ. **149**, 283 (1967).

[76] R. Pound, G. Rebka: "Apparent weight of photons", Phys. Rev. Lett., **4**, 337 (1960), e successive versioni più sofisticate.

Su (d): La CMB fu scoperta in modo quasi casuale nel 1964 da A. Penzias (1933-) e R. Wilson (1936-) durante la calibrazione di un'antenna a microonde progettata per le comunicazioni satellitari, e mentre esperimenti mirati alla sua rivelazione erano già in via di realizzazione. La radiazione risultò avere un altissimo (entro ca. $10^{-5}$) grado di isotropia ed un preciso (entro ca. $10^{-4}$) spettro di equilibrio termico intorno alla temperatura di 2,728 K. La scoperta della CMB (per la quale Penzias e Wilson ricevettero il Nobel) fu *veramente fondamentale*, segnando in pratica la fine della lunga diatriba tra i sostenitori della teoria del big-bang di Friedmann-Lemaître e quelli della "teoria dello stato stazionario", proposta da Hoyle, Gold (Thomas, 1920-2004) e Bondi (Herman, 1919-2005). [77]

Su (e): Il ritardo dei segnali elettromagnetici fu osservato nel 1960 utilizzando intensi impulsi radar inviati verso Venere o Mercurio quando questi pianeti si trovavano al bordo del Sole, e da essi riflessi.

Su (f): La formazione di immagini multiple, effetto noto come "di lente gravitazionale", fu osservata per la prima volta agli inizi degli anni '80 e riguardò una quasar situata appunto dietro una galassia abbastanza trasparente. #

Ulteriori importanti conferme sperimentali, sulle quali sorvoliamo, si sono registrate negli ultimi 2-3 decenni. Alcune necessarie implicazioni della teoria sono invece ancora oggetto di ricerca, prima tra tutte quella della possibile rivelazione della radiazione di gravità che essa prevede; e ciò sia sotto la forma di tenuissime vibrazioni meccaniche di antenne massive (sopraraffreddate) che come perturbazioni della metrica rivelabili da grandi (parecchi km) o grandissime (intersatellitari, attualmente in progetto) antenne interferometriche. Come naturale, il grande problema delle antenne terrestri è quello del rumore che maschera i possibili segnali utili, problema al quale si cerca di ovviare mediante tecniche di coincidenza remota. Si può infine ricordare che recenti osservazioni della pulsar binaria PSR 1913+16 stabiliscono che questo sistema perde energia ad un tasso corrispondente con buona approssimazione all'emissione di onde di gravità secondo il modello relativistico.

Nel concludere questa breve sintesi, non possiamo non rimarcare che la teoria geometrica della gravitazione di Einstein (o secondo alcuni, di Einstein-Hilbert) è talmente attraente e profonda, e sembra andare così al cuore della fisica, che una sua ormai estremamente improbabile falsificazione – beninteso nella scala *macroscopica* alla quale essa deve essere riferita – costituirebbe senza dubbio un drammatico vulnus per la filosofia naturale contemporanea.

---

[77] Alcune importanti divergenze tra le iniziali stime di Hubble sull'età dell'universo (ca 1,8 miliardi di anni) e quelle decisamente maggiori che dovevano presupporsi sulla base di altre e indipendenti valutazioni furono gradualmente appianate dalle successive ricerche di W. Baade (1952, ca 3,6 miliardi di anni) e soprattutto di A. Sandage, che intorno al 1960 portò l'età in questione ad un valore compreso tra i 10 e i 20 miliardi di anni (le stime attuali si aggirano sui 14-15 miliardi di anni). Questi valori corrispondono a dimensioni dell'universo potenzialmente visibile di circa $10^{26}$ m.